\documentclass[proof]{WileyASNA-v1}

\articletype{Article Type}%

\received{1 October 2020}
\revised{12 October 2020}
\accepted{25 October 2020}

\begin{document}

\title{Stability of multi-state configurations of fuzzy dark matter}

\author{Francisco S. Guzm\'an}


\address{\orgdiv{Instituto de F\'{\i}sica y Matem\'{a}ticas, Universidad
              Michoacana de San Nicol\'as de Hidalgo. Edificio C-3, Cd.
              Universitaria, 58040 Morelia, Michoac\'{a}n,
              M\'{e}xico.}}

\corres{\email{francisco.s.guzman@umich.mx}}

\abstract{We study the stability properties of multi-state configurations of the Schr\"odinger-Poisson system without self-interaction, with monopolar and first dipolar components $(1,0,0)$+$(2,1,0)$. We show these configurations studied are stable using numerical simulations, and using criteria of stationarity, unitarity and time dependence consistency. The study covers a range of states with  monopolar to dipolar mass ratio between 47 and 0.17. The astrophysical implication of this result is that this type of configurations is at least stable and  can be considered physically sound in multi-state ultralight bosonic dark matter.}


\maketitle

\section{Introduction}

One of the dark matter candidates that has attracted considerable attention recently is an ultralight boson. The model started in the cosmological context, where a scalar field was proposed to be the dark matter with the aim of explaining the abundance of small structures that cold-dark-matter simulations predicted to be higher than observed. The mass power spectrum could be fitted with a single parameter, namely the mass of the scalar field, which was restricted to be ultralight, with mass of order $\sim 10^{-22}eV/c^2$, \citep{TonaLuis2001}, \citep{Hlozek:2014lca}, \citep{Hui:2016ltb}. In order to deal with the phenomenology of structures in the non-linear regime, after the turn-around point, it was necessary to adapt the Einstein-Klein-Gordon (EKG) equations to the low energy and weak gravitational field regimes, which leads to the Schr\"odinger-Poisson (SP) system of equations. In this regime the model has been studied from different angles, including stationary regimes like  galactic rotation curves \citep{FluidDarkMatter}, or the Gravitational cooling as a relaxation process in structure formation \citep{GuzmanUrena2006}.

Studies that have boosted the model involve  structure formation studies, and the comparison with the process within the CDM paradigm \citep{Schive:2014dra,Mocz:2017wlg,PhysRevD.98.043509}. These analyses indicate that stationary solutions of the SP system are attractors not only in simple scenarios \citep{BernalGuzman2006A,PhysRevD.94.043513}, but also in structure formation conditions \citep{PhysRevLett.113.261302}. State of the art simulations of structures assuming ultralight bosonic dark matter, involve the processes of interaction of galactic halos with luminous matter, including the scenario of star formation \citep{mocz19,PhysRevD.101.083518}.

All these advances are based on the assumption that  the bosons are in a single coherent state. Further generalizations include excited states considered to be surrounding ground state galactic cores \citep{pozo2020wave} and recently multi-state configuration, not-only a wave-like sea of matter in excited states, but actual stationary multi-state solutions of the SP system have been proposed to be a possible explanation of the Vast Polar Structre problem \citep{sopez2020scalar}, that indicates that in some galaxies, including the Milky Way, M31 and Centaurus A, satellite galaxies distribute anisotropically \citep{Conn_2013,ibata2013vast,Pawlowski:2019bar}.

Multi-state configurations of the SP system were proposed since the sixties \citep{RuffiniBonazzolla},  more recently  the case of spherical modes has been studied in detail \citep{UrenaBernal2010} and a detailed analysis of configurations that include axially symmetric multi-state boson star configurations has been already developed \citep{GuzmanUrena2020}. With this knowledge at hand and the possible relevance of excited state or multi-state configurations having a possible role in dark matter scenarios, in this paper we explore the stability of two-state configurations, including the monopolar (1,0,0) and the first dipolar state (2,1,0).

The paper is organized as follows. In Section \ref{sec:configs} we describe the elements used to construct equilibrium configurations. In section \ref{sec:stability} we define the criteria to determine stability, whereas in section \ref{sec:comments} we draw some conclusions.

\section{Configurations}
\label{sec:configs}

Equilibrium multi-state configurations can be constructed following the recipe in \citep{GuzmanUrena2020}, which can be summarized as follows. The dimensionless SP system of equations for a multi-state configurations can be written as

\begin{eqnarray}
i\frac{\partial \Psi_{nlm}}{\partial t} &=& -\frac{1}{2}\nabla^2 \Psi_{nlm} + V\Psi_{nlm},\label{eq:schro}\\
\nabla^2 V &=& \sum_{nlm} |\Psi_{nlm}|^2,\label{eq:Poisson}
\end{eqnarray}

\noindent where the wave function $\Psi_{nlm}(t,{\bf x})$ is an order parameter describing the dynamics of the macroscopic interpretation of the bosonic gas in the state $nlm$, with $|\Psi_{nlm}|^2$  the mass density in such state. Finally,  $V(t,{\bf x})$ is the gravitational potential acting as a trap for the bosonic cloud, sourced by the addition of the mass densities of each state. Notice that there are no crossed terms among different states in Poisson equation. In principle one would expect  a general state to be a linear combination of particular states, for example $\Psi=\sum_{nlm}\Psi_{nlm}$, then the total mass density would be $|\Psi|^2=\left(\sum_{nlm}\Psi_{nlm}^{*}\right) \left(\sum_{n'l'm'}\Psi_{n'l'm'}\right)$. However, the SP system (\ref{eq:Poisson}) does not contain the crossed terms because it is derived as the low energy and weak gravitational field limits of the Einstein-Klein-Gordon equations. In such frame, the normal ordering condition requiring  vacuum to have a zero density  energy density, implies all the crossed terms vanish. A detailed demonstration of this condition can be seen in \citep{RuffiniBonazzolla,UrenaBernal2010}.

The system above has stationary solutions when  the wave functions are assumed to have a harmonic time-dependence.  This condition combined with the use of spherical coordinates, and an expression of each wave function $\Psi_{nlm}$ on the spherical harmonic basis, one proposes the wave functions to be of the form

\begin{equation}
\Psi_{nlm}(t,{\bf x}) = \sqrt{4\pi} e^{-i\gamma_{nlm}}r^l \psi_{nlm}(r)Y_{lm}(\theta,\phi)
\label{eq:ansatz}
\end{equation}

\noindent where $\psi_{nlm}(r)$ are radial dependent wave functions and $\gamma_{nlm}$ are frequencies of oscillation of each state. Harmonic time dependence implies that the source of Poisson equation in (\ref{eq:Poisson}) is time independent, and thus $V$ should be time independent as well. It also reduces the general Schr\"odinger equation for each state (\ref{eq:schro}) to a stationary Schr\"odinger equation. The resulting set of stationary Schr\"odinger equations is solved as a set of eigenvalue problems, one for each $\psi_{nlm}$ obeying boundary conditions. The boundary conditions consist in smoothness at the origin, which fixes the values of the wave functions $\psi_{nlm}(0)$ to finite constants, and their spatial derivatives $\psi_{nlm}'(0)$ to zero. On the other hand, at infinity isolation conditions are ompised as follows. The radial coordinate in practice covers a finite domain $r\in [0,r_{max}]$, and it is at $r_{max}$ where the wave functions  and their spatial derivatives are required to be small, with $|\psi_{nlm}(r_{max})|<\epsilon$ and $|\psi_{nlm}'(r_{max})|<\epsilon$, with $0<\epsilon\ll 1$ a small tolerance value. Imposing these boundary conditions defines a Sturm-Liuoville problem, where the eigenvalues are the frequencies $\gamma_{nlm}$ of each state,  that we solve using a shooting method \citep{GuzmanUrena2020}.

In this paper we focus on a particular type of mixed configurations that appear to be of astrophysical interest, related to a possible explanation of the Vast Polar Structure problem \cite{sopez2020scalar}. These particular configurations are mixd state solutions that combine the first spherical and the first non-spherical modes $(1,0,0)$ and $(2,1,0)$ respectively. In order to investigate how reliable these configurations can be as galactic dark matter halos, it is very important to determine whether they are stable or not.

\begin{figure}[htb]
\centering
\includegraphics[width=7cm]{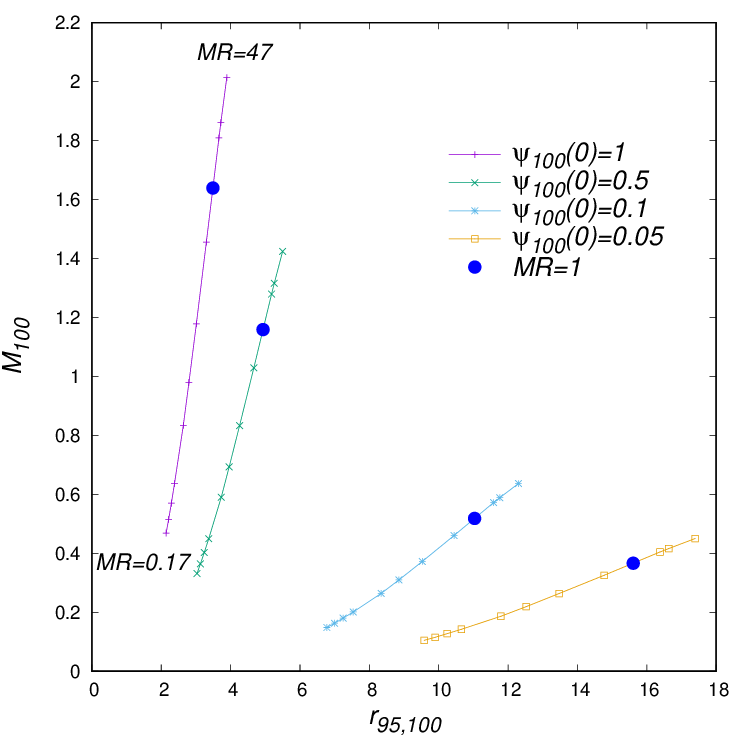} 
\caption{Families of solutions constructed with four values of $\psi_{100}(0)$ and many values of $\psi_{210}(0)$, represented in a mass versus radius diagram. We chose the mass to be that of the spherical contribution and the radius $r_{95}$ of the spherical density distribution. We label these quantities the $M_{100}$ and $r_{95,100}$ respectively. Indicated with filled circles are the configurations with $MR=M_{100}/M_{210}=1$, where $M_{100}$ is the mass of the monopolar contribution and $M_{210}$ that of the dipolar contribution. Also indicated in the first of the families, the configurations with the highest and smallest value of $MR$ in our analysis.}
\label{fig:eqconfigs}
\end{figure}

In the case of the spherically symmetric equilibrium solutions \citep{GuzmanUrena2004}, the family of equilibrium configurations depends on one parameter, namely the central value of the unique wave function $\psi_{100}$. In the case of a configuration with two states $(1,0,0)$ and $(2,1,0)$, there are two parameters that characterize the stationary solutions, the central value of the two involved wave functions $\psi_{100}(0)$ and $\psi_{210}(0)$. In Figure \ref{fig:eqconfigs} we show a sample of families of equilibrium configurations with various central values of the spherical wave function $\psi_{100}(0)$, and for each of these a number of  central values for the dipolar contribution $\psi_{210}(0)$.

Each value of $\psi_{100}(0)$  defines a family of solutions in terms of the central value of $\psi_{210}$. Notice that once the value of $\psi_{100}(0)$ has ben chosen, and its associated  family constructed, the others can be reproduced by using the scaling relations of the SP system for the wave function $\Psi_{210}$. That is, for an arbitrary $\lambda$, if the wave function scales like $\psi = \lambda^2 \hat{\psi}$,  then mass and length scale as  $M =\lambda \hat{M},~r=\hat{r}/\lambda$ respectively \citep{GuzmanUrena2004}. This means that once we construct the family for $\psi_{100}(0)=1$, it is possible to generate the other families with this scaling, something we verified with the four families in Figure 1, where the families  corresponding to $\psi_{100}(0)=0.5,~0.1,~0.05$ can be obtained by rescaling the family of $\psi_{100}(0)=1$ with $\lambda=
\sqrt{.5},\sqrt{.1},\sqrt{0.05}$ respectively. In this way, all the families in between these four particular cases can also be generated.

One of the important properties of these 2-state configurations is the mass ratio between states. If the mass of the (1,0,0) state is defined as $M_{100}=\int |\Psi_{100}|^2d^3 x$, and the mass of the state (2,1,0) is $M_{210}=\int |\Psi_{210}|^2d^3 x$, we define the mass ratio between states $MR=M_{100}/M_{210}$. If $MR>1$ we say the configuration is dominated by the monopole contribution, whereas if $MR<1$ we say the configuration is dipole dominated.

The configurations in  Figure \ref{fig:eqconfigs} contain configurations ranging from very monopole dominated configurations with $MR=47$ to very dipole dominated configurations with  $MR=0.17$. Whether or not they can be used to model dark matter halos will depend on the comparison with observations of each galaxy hosting satellites. However, in this paper we focus  on the stability properties of  these solutions only.

\section{Stability}
\label{sec:stability}

The stability can be studied with different approaches, for example using perturbation theory or by evolving the configurations and determine they are long lived. Here we follow the second approach and 
investigate the stability by following the evolution in full 3D of the configurations. 
For this we inject the wave functions $\Psi_{100}$ and $\Psi_{210}$ of equilibrium configurations into the 3D numerical domain. 
We then evolve these equilibrium configurations by solving the fully time-dependent equations in a 3D cubic domain described with Cartesian coordinates $[-x_{max},x_{max}]\times [-y_{max},y_{max}]\times [-z_{max},z_{max}]$, using the finite differences approximation of spatial derivatives in Eqs. (\ref{eq:schro}-\ref{eq:Poisson}), and the method of lines for the evolution that uses a third order Runge-Kutta integrator. Poisson equation is used with a Successive Overrelaxation method with optimal acceleration parameter. We use fixed mesh refinement in order to achieve higher precision at the center of the domain where the configuration is sitting, and at the same time boundary conditions at a considerable distance, being located at five times $r_{95}$ the radius containing 95\% of the mass of the configuration. In order to make more evident the effect of numerical perturbations, we use a rather small resolution of $\Delta x=\Delta y=\Delta z=0.2$ in the refined level, which is approximately $r_{95}/20$. At the same time, in order to avoid reflection from the boundaries, we use a sponge consisting of an imaginary potential \citep{GuzmanUrena2004}, absorbing the density approaching the boundaries, that we locate at a radius of 0.9$x_{max}$.
Due to numerical and truncation errors, the configuration at initial time contains already a perturbation, which produces oscillations in the configurations. This method has been used in similar systems including Boson Stars  to analyze stability and quasinormal oscillation modes (see e.g. \cite{Guzman2004}).

The criteria used to determine whether or not a configuration is stable involves various aspects of the evolution, including stationarity, unitarity of the evolution and consistency of the eigenfrequency of each state with the frequency of the wave functions during the evolution.

Stationarity refers to the time independence of the density of each state composing the configuration. In theory both $|\Psi_{100}|^2$ and $|\Psi_{210}|^2$ densities should remain nearly time independent during the evolution. In practice this is not as exact. In previous analysis involving single state configurations, for example ground state configurations \citep{GuzmanUrena2004,GuzmanUrena2006} show a clear stationarity even when evolved in a 3D cartesian domain with low resolution \citep{GuzmanAvilez2018}, and mixed state spherical configurations can be seen stationary with spherical codes \citep{UrenaBernal2010}, and show bigger oscillations when resolution is low, for instance in full 3D \citep{GuzmanAvilez2018}. This is a first criterion to be fulfilled, if evidently the densities change shape or disperse away, the configuration is clearly stationary. This is a first test applied to the sample of configurations studied.

Unitarity is another characteristic of a numerical evolution that is not exact. In theory $M_{100}=\int |\Psi_{100}|^2 d^3 x$ and $M_{100}=\int |\Psi_{100}|^2 d^3 x$, integrated in the whole numerical domain, should remain time-independent, however, the numerical methods involved, and the dissipation that carry on with them, destroy the exact unitarity of the evolution. Then in practice one preserves the number of particles in each state only approximately, within a certain percentage error. At this respect we measured the violation of unitarity for a representative sample of configurations.

The third criterion is Frequency of oscillations. In order to consider that the configuration that is being evolved is the correct one, this is the most solid test, which is related to the frequency of the wave functions $\Psi_{100}$ and $\Psi_{210}$ during the evolution. In theory it should coincide with the eigenvalue obtained when solving the Strum-Liouville problem, that is $\gamma_{100}$ and $\gamma_{210}$, as seen in the ansatz (\ref{eq:ansatz}). What is done to investigate whether this is the case or not, is to calculate the peak frequencies from a   Fourier Transforms of the maximum of the wave functions $\Psi_{100},\Psi_{210}$ after a long term evolution.

\begin{figure}[htb]
 \centering
\includegraphics[width=4cm]{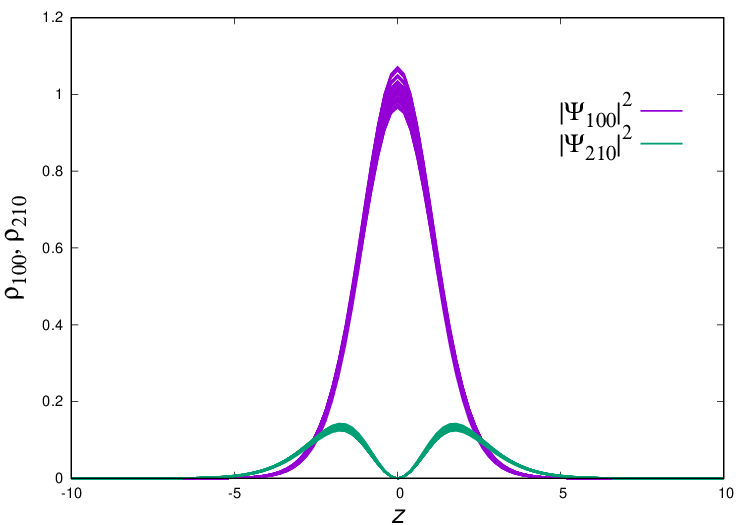} 
\includegraphics[width=4cm]{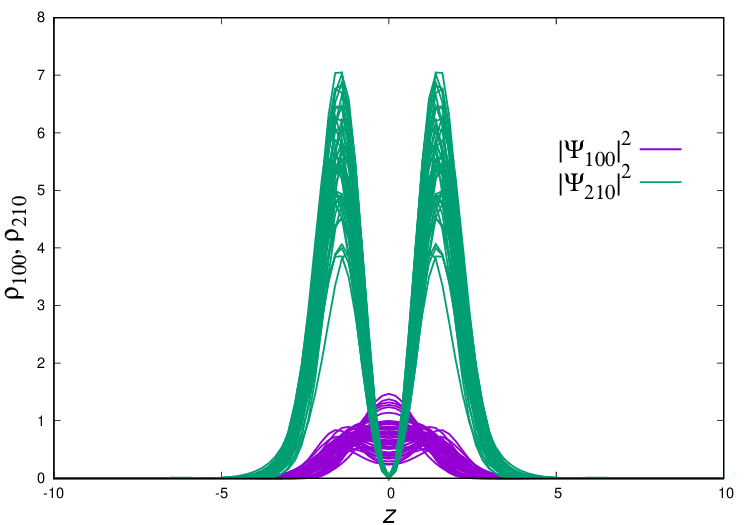} 
\includegraphics[width=4cm]{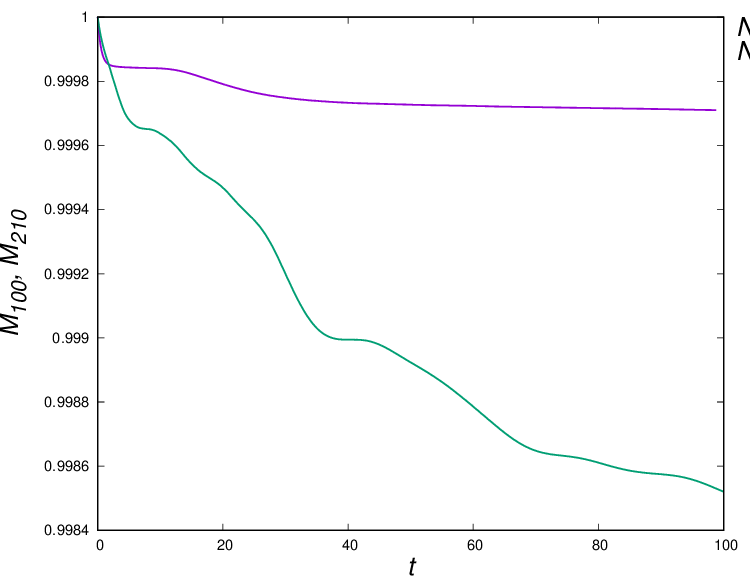} 
\includegraphics[width=4cm]{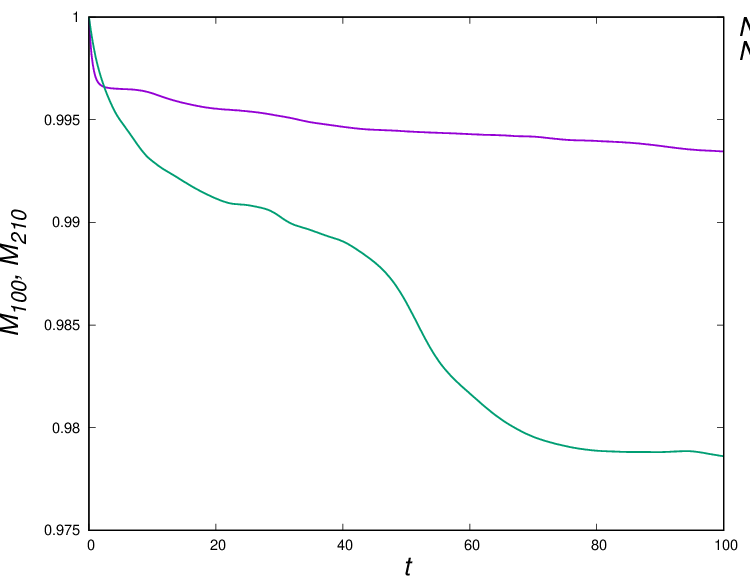} 
\includegraphics[width=4cm]{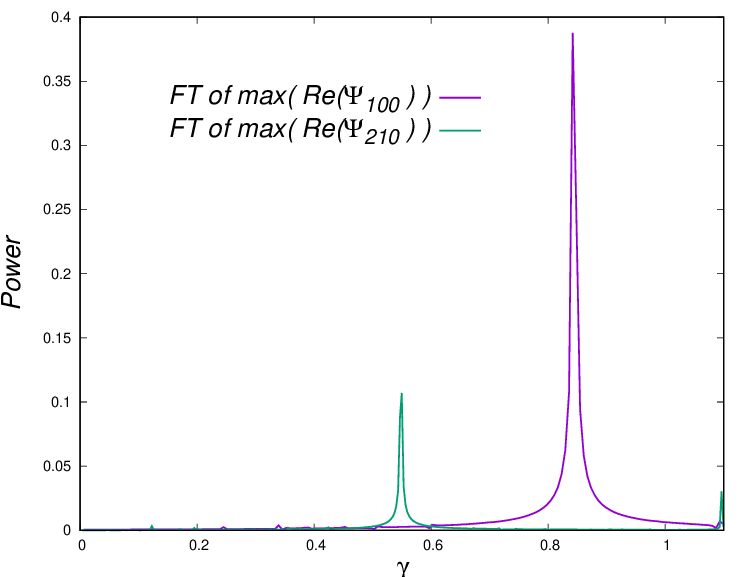} 
\includegraphics[width=4cm]{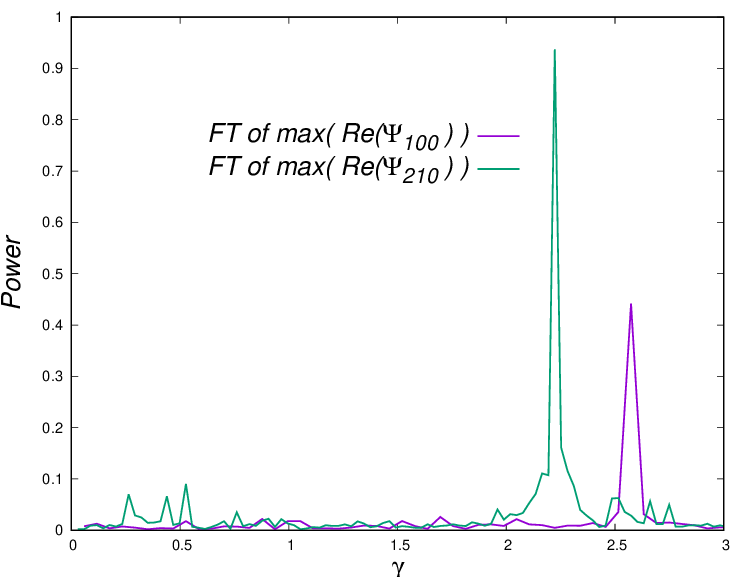} 
\caption{Diagnostics for two configurations, respectively the ones at the extremes for the family with $\psi_{100}(0)=1$. These are, the most monopole dominating one with $\psi_{210}(0)=0.2$ at the left, and the most dipole dominating configuration with $\psi_{210}(0)=1.5$ at the right. In the first row we  show snapshots of $|\Psi_{100}|^2$ and $|\Psi_{210}|^2$, which indicate the densities are not exactly stationary, but oscillate. In the middle we show the amount of mass lost during the evolution, normalized with the mass of each state at initial time; in the ideal case, a perfectly unitary evolution would indicate no mass is lost at all, whereas in a real simulation some mass is lost. The first configuration loses less than 1\% whereas the second configuration loses about 2\%. The time domain is in code units, and covers dozens of oscillations of $\Psi_{100}$. Finally, in the third row we show the Fourier Transform of the maximum of $\Psi_{100}$ and $\Psi_{210}$ and show the peak frequencies correspond to the frequencies calculated when solving the eigenvalue problem for stationary configurations.} 
\label{fig:stable}
\end{figure}

In Figure \ref{fig:stable} we show these three criteria for the most extreme of the configurations in the family of equilibrium configurations with $\psi_{100}(0)=1$, which, as mentioned before, can be converted into any other family. By extreme configurations we refer to the most monopole dominated  with $MR=47$ and the most dipole dominated $MR=0.17$ configurations. The first criterion is easily sorted out in the first case, and with more difficulty by the second configuration. The number of particles is preserved nearly in the same proportion in the two cases, whereas the strong test, concerning the oscillation frequencies shows very similar accuracy.

In Table \ref{tab:freqs} we indicate the frequencies from the eigenvalue problem, and the deviation from these values obtained after the evolution was carried out, for a sample of the two extreme configurations and three intermediate ones.

\begin{table}
\begin{tabular}{ccccc}\hline
$\phi_{210}(0)$ & $\gamma_{100}^{st}$ & $\gamma_{210}^{st}$ &$\Delta \gamma_{100}$ (\%) &$\Delta \gamma_{210}$ (\%)  \\\hline
0.2			& 0.8355 & 0.5381 	&	0.77	& 0.65\\
0.5			& 1.2554 & 0.9280	&	1.49	& 2.01 \\
0.9			& 1.8011 & 1.4255	&	0.92	& 2.05 \\
1.201		& 2.1736 & 1.7603	&	0.3	& 1.83\\
1.504		& 2.5238 & 2.0728	&	0.95	& 1.75 \\\hline
\end{tabular}
\caption{Frequencies obtained by solving the Sturm-Liouville problem for a selection of stationary configurations with $\psi_{100}(0)=1$, scalable to other central values of the monopolar wave function's central value, covering from the most monopolar configuration to the most dipolar configuration in Figure \ref{fig:eqconfigs}. We present the frequency of the two states $\gamma^{st}_{100}$ and $\gamma^{st}_{210}$. We also show the percentage error of the frequency of oscillation of the wave functions $\Psi_{100}$ and $\Psi_{210}$, calculated with a Fourier Transform after the system has evolved. The error in all cases is smaller than  1\% in the spherical mode, and of order  2\% in the dipolar mode.}
\label{tab:freqs}
\end{table}

\section{Conclusions}
\label{sec:comments}

In this paper we have used the criteria of stationarity, unitarity and harmonic time dependency, to determine the stability of mixed state solutions with states $(1,0,0)$ and $(2,1,0)$, of the Schr\"odinger-Poisson system of equations.

The parameter space studied covers a wide range of state domination between monopole and dipole terms. The parametrization in terms of $MR$  covers an enormous range of configurations with astrophysical applications within the ultralight dark matter model scenario, because the mass of the boson is the free parameter that fixes the spatial scale.

The analysis presented here, allows  the possibility to consider multi state dark matter scenarios, which can help explaining the wavy behavior of dark matter near galactic cores under the BEC dark matter hypothesis.

\bibliography{Stability}%

\end{document}